\documentclass{article}
\usepackage{graphicx}
\usepackage{hyperref}
\usepackage{todonotes}
\usepackage[noblocks]{authblk}

\begin{document}

\title{Charting the Future of Scholarly Knowledge with AI: A Community Perspective}

\author[1, 2]{Azanzi Jiomekong}%\email{jiomekong@tib.eu} 
\author[4]{Hande Küçük McGinty}%\email{hande@ksu.edu} 
\author[5]{Keith G. Mills}%\email{Keith.Mills@lsu.edu} 
\author[1]{Allard Oelen}%\email{allard.oelen@tib.eu} 
\author[6]{Enayat Rajabi}%\email{Enayat_Rajabi@cbu.ca} 
\author[7]{Harry McElroy}%\email{hmcelroy@performigence.com} 
\author[8]{Antrea Christou}%\email{christou.2@wright.edu} 
\author[8]{Anmol Saini}%\email{saini.25@wright.edu} 
\author[9]{Janice Anta Zebaze} %
\author[10]{Hannah Kim}%\email{hannah.kim0007@temple.edu} 
\author[11]{Anna M. Jacyszyn}%\email{anna.jacyszyn@fiz-karlsruhe.de} 
\author[1]{Gollam Rabby}
\author[1]{Dirk Betz}
\author[1]{Claudia Biniossek}
\author[12]{Sanju Tiwari}
\author[1, 3]{Sören Auer} 

\affil[1]{TIB Leibniz Information Centre for Science and Technology \\ \texttt{\{jiomekong, allard.oelen, auer\}@tib.eu}}%, Hannover, Germany}
\affil[2]{Department of Computer Science, University of Yaounde 1} %, Cameroon}
\affil[3]{L3S Research Center, Leibniz University of Hannover} %, Germany}
\affil[4]{Department of Computer Science, Kansas State University \\ \texttt{hande@ksu.edu}} %, Manhattan KS}
\affil[5]{School of EECS, Louisiana State University \\4 \texttt{Keith.Mills@lsu.edu}}
\affil[6]{Management Science Department, Cape Breton University \\ \texttt{Enayat\_Rajabi@cbu.ca}} %, Sydney, NS, Canada}
\affil[7]{Department of Development and Research, Performigence \\ \texttt{hmcelroy@performigence.com}} % Corporation, Los Angeles, CA, USA}
\affil[8]{Department of Engineering and Computer Science, Wright State University \texttt{\{christou.2, saini\}@wright.edu}} %, Dayton, OH, USA}
\affil[9]{Department of Physics, University of Yaounde 1 \\ \texttt{antazebaze@gmail.com}} %, Cameroon}
\affil[10]{Department of Biology, Temple University \\ \texttt{hannah@kim0007@temple.edu}} %, Philadelphia, PA, USA}
\affil[11]{FIZ Karlsruhe, Leibniz Institute for Information Infrastructure \\ \texttt{anna.jacyszyn@fiz-karlsruhe.de}} %, Hermann-von-Helmholtz-Platz 1, 76344 Eggenstein-Leopoldshafen, Germany}
\affil[12]{Sharda University, Delhi-NCR, India}

%\affil[7]{}

\maketitle

\newpage

\maketitle

\begin{abstract}
Despite the growing availability of tools designed to support scholarly knowledge extraction and organization, many researchers still rely on manual methods, sometimes due to unfamiliarity with existing technologies or limited access to domain-adapted solutions. Meanwhile, the rapid increase in scholarly publications across disciplines has made it increasingly difficult to stay current, further underscoring the need for scalable, AI-enabled approaches to structuring and synthesizing scholarly knowledge. Various research communities have begun addressing this challenge independently, developing tools and frameworks aimed at building reliable, dynamic, and queryable scholarly knowledge bases. However, limited interaction across these communities has hindered the exchange of methods, models, and best practices, slowing progress toward more integrated solutions. This manuscript identifies ways to foster cross-disciplinary dialogue, identify shared challenges, categorize new collaboration and shape future research directions in scholarly knowledge and organization.

\textbf{keywords}{AI for Scholarly Communication, Scholarly Knowledge Extraction, Scholarly Knowledge Organization, Scholarly Knowledge Generation, Scholarly Knowledge Use, AI- Ethics in Scholarly Communication}
\end{abstract}

\section{Introduction}
Scholarly work and communication encompass the entire system in which research and creative works are created, evaluated for quality, disseminated to the academic community and beyond, used, and preserved for future use. It includes formal publications, such as journal articles and books, as well as informal sharing through preprints, conference presentations, data sharing, and broader engagement with scholarly works and research outputs. Scholarly knowledge serves as the primary engine of progress, shaping our world and guiding our collective future. It forms the backbone of technological advancement, public health systems, and sustainable environmental practices. Obtained through rigorous methods of observation, experimentation, and validation, it is a reliable resource that helps societies solve complex problems and improve the quality of life by achieving sustainable development goals (SDGs)~\cite{UNSDGs2015}.

To be truly useful, scholarly knowledge must first be systematically extracted and organized. However, the scholarly community of today faces the problem of an overload of scientific papers in their respective domains. There is an increasing number of papers published every year (currently, ~3 million), in addition to more than 200 million papers that have already been published . This gives rise to the research question: ``How can we provide a reliable and living scholarly knowledge base that empowers researchers to query, synthesize, and analyze the vast body of scholarly knowledge?"~\cite{JiomekongWWWConf2024}. To solve this problem, researchers are increasingly using AI to extract, organize~\cite{JiomekongWWWConf2024,jiomekongORKG2024,GreinerSemMathFormulae2023,surveyIEScienPaperZara2018,KEReviewAbdul2020}, analyze, and synthesize knowledge~\cite{ImprovingAccessAuer2020, lu2024ai, gottweis2025towards, eger2025transformingsciencelargelanguage, van2023artificial, bolanos2024ALiteReviews,ManghiOpenAire2012,manghiOpenAire2026} from scholarly literature.

Researchers have contributed to the development of AI-based systems for scholarly communication. Several recent studies described these studies by considering several dimensions: 
\begin{itemize}
    \item AI systems for literature review~\cite{bolanos2024ALiteReviews, whitfield2023elicit, stephen2011library, o2014techniques, bernard2025using, heikkila2024human},
    \item AI models, datasets, and tools for different aspects of the research life cycle~\cite{eger2025transformingsciencelargelanguage, CORE-GPT, jhajj2024use, brack2022cross, korinek2023generative, fisher2016natural},
    \item The exploration of the capability of Large Language Models (LLMs) in generating novel research ideas based on information from research papers~\cite{eger2025transformingsciencelargelanguage, bolanos2024ALiteReviews}.
\end{itemize}

The research community working on AI for Scholarly communication (composed of scholarly AI developers and users) has operated more or less independently, with limited exchange of ideas, methodologies, methods, theories, etc. The inaugural edition of the \href{https://sites.google.com/view/ai4sc/edition/ai4sc-AAAI2025}{AAAI Bridge on AI for Scholarly Communication (AI4SC)} at \href{https://aaai.org/conference/aaai/aaai-25/aaai-25-bridge-list/#bp02}{AAAI 2025} brought together a diverse audience, including students, domain researchers, practitioners, and AI experts, regardless of their current use of AI technologies. The goal was to foster cross-disciplinary dialogue, identify shared challenges, catalyze new collaborations, and shape future research directions in scholarly knowledge extraction, organization, and use. This event spanned two days, and this paper aims to report its findings, results of discussions, and suggested future directions, as well as the challenges discussed.

\begin{figure}
    \centering
    \includegraphics[scale=0.4]{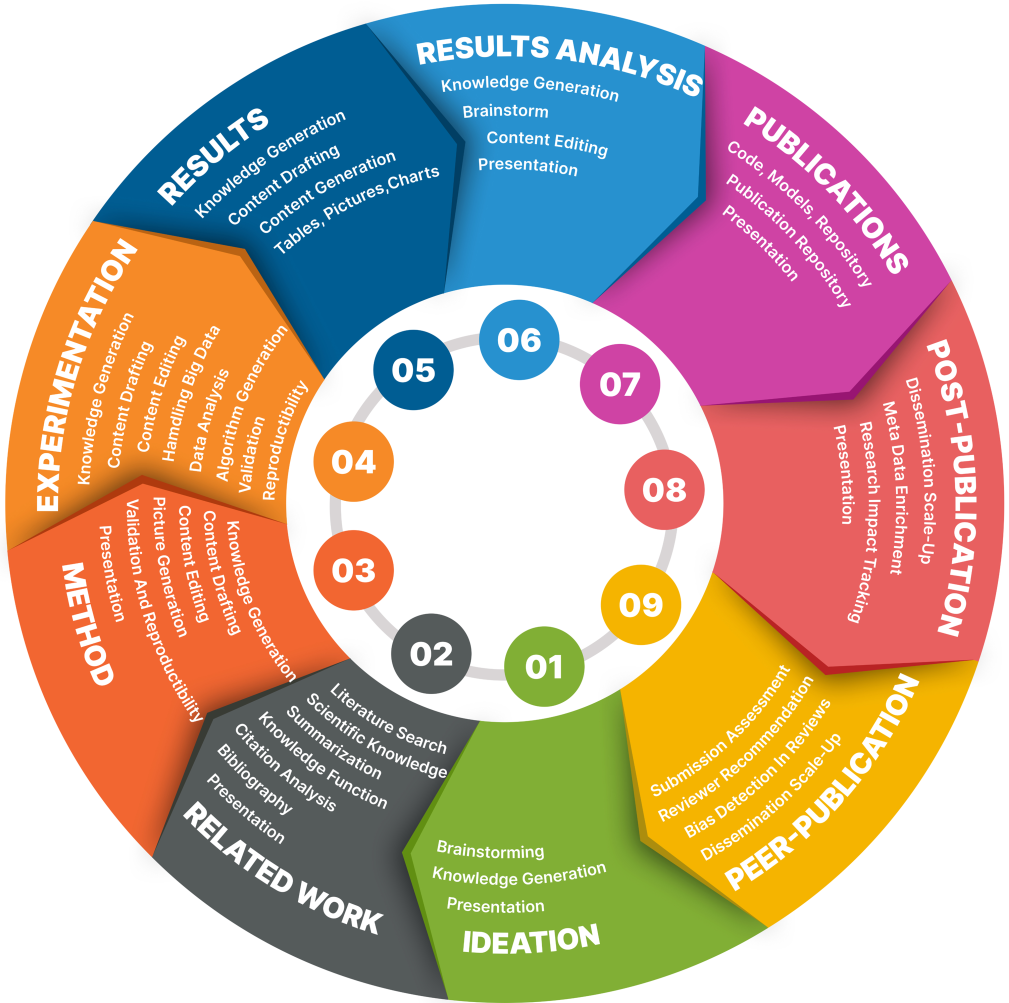}
    \caption{Illustration of the research life cycle. For a given problem, scientists first brainstorm potential solutions, incorporating and improving on existing literature in order to propose a new method. They conduct experiments to verify the efficacy of their approach, and analyze the results. Once satisfactory findings have been attained, the research is formalized into a publication and subjected to the peer-review process. This adds to the existing literature body and provides new avenues for brainstorming.}
    \label{fig:researchLifeCycle}
\end{figure}

Through a combination of technical talks, collaborative workshops (such as world café sessions), and thematic discussions, participants explored a range of pressing issues at the intersection of AI and scholarly knowledge production and use. Central themes included the \textbf{classification of scholarly knowledge}, where participants examined challenges and advances in categorizing and structuring research across diverse domains. Discussions on \textbf{knowledge acquisition and organization} focused on the development of tools and frameworks for extracting, linking, and maintaining scholarly knowledge at scale. The \textbf{impacts of Generative AI (GenAI) and Artificial General Intelligence (AGI)} were also a key topic, highlighting both the opportunities and risks associated with these technologies throughout the research lifecycle (see Fig. \ref{fig:researchLifeCycle}). \textbf{Community engagement and participation} emerged as another important theme, emphasizing strategies to foster inclusivity, transparency, and interdisciplinary collaboration. Participants also reflected on \textbf{knowledge utilization}, particularly how to empower researchers with actionable insights and decision-support systems derived from structured knowledge. Finally, \textbf{ethical considerations} were at the forefront, including concerns about bias, provenance, reproducibility, and the responsible use of AI in scholarly discourse. The following sections present the outcome of these discussions and the opportunities and challenges.

\section{Classification of AI Systems in Scholarly Communication}
From narrow assistive tools to more autonomous generative systems, AI is reshaping how scholarly knowledge is created, evaluated, disseminated, and consumed~\cite{eger2025transformingsciencelargelanguage,bolanos2024ALiteReviews}, and in doing so, playing a transformative role across the entire research lifecycle (see \autoref{fig:researchLifeCycle}). Classifying AI systems by the stages of this lifecycle helps to clarify their capabilities, benefits, and challenges in transforming scholarly knowledge creation, dissemination, and use. This classification will help researchers, publishers, institutions, and policy-makers to select appropriate tools and anticipate their limitations.

\begin{figure}
    \centering
    \includegraphics[scale=0.17, angle=90]{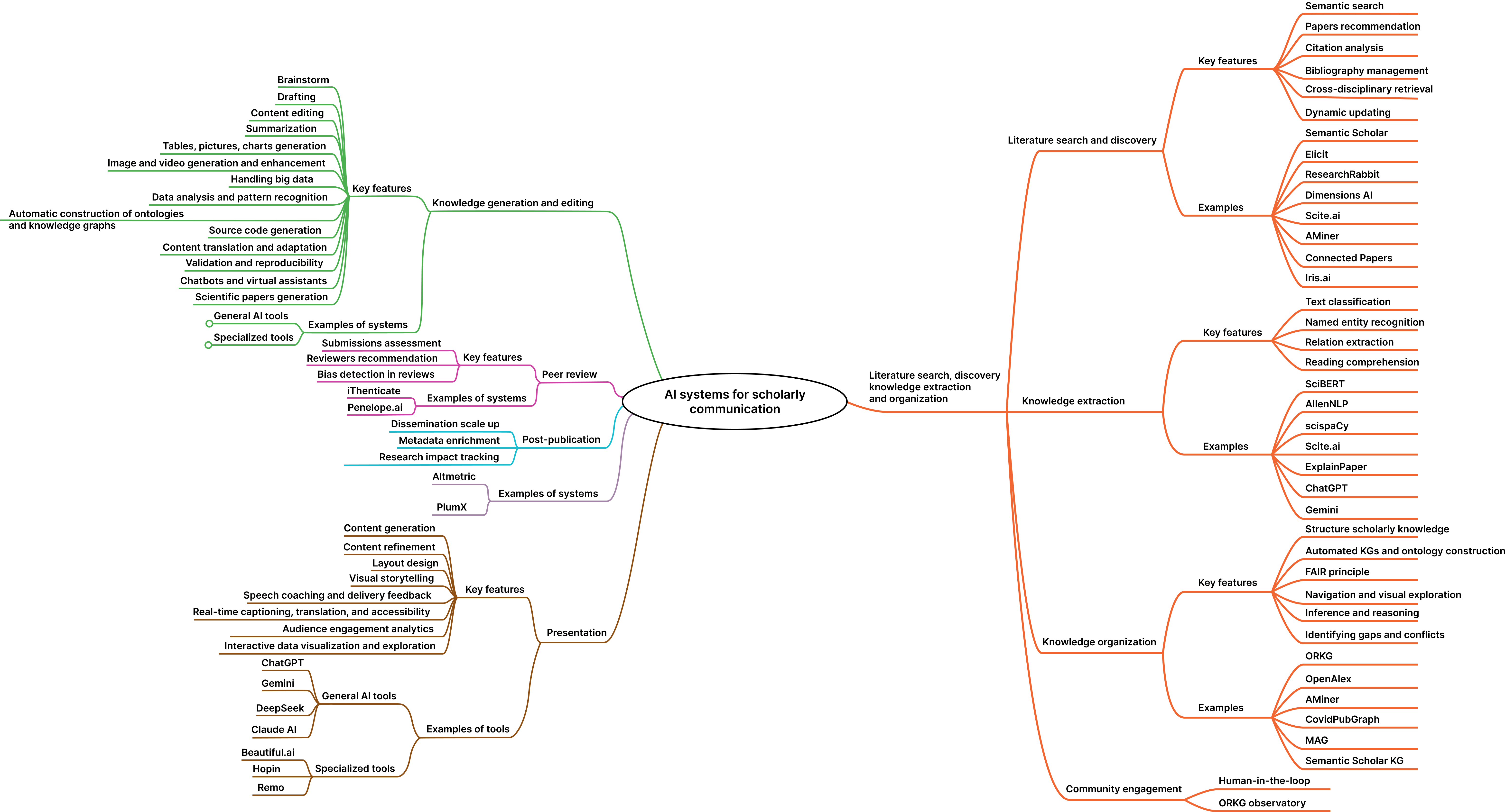}
    \caption{Holistic view of the classification of artificial intelligence for scholarly communication}
    \label{fig:hollisticViewAI4SC}
\end{figure}

Generally, AI systems for scholarly communication can be roughly classified as follows:

\begin{itemize}
    \item Systems for literature discovery, scholarly knowledge extraction and organization;
    \item Systems for scholarly knowledge generation;
    \item Systems for peer review, publication, post-publication;
    \item Systems for slide presentation.
\end{itemize}

\autoref{fig:hollisticViewAI4SC} presents a holistic view of AI systems for scholarly communication. The components presented in this figure are described below.

\subsection{AI Systems for Literature Search, Discovery, knowledge extraction, and organization}
Scholarly communication is recorded, indexed, and disseminated in scholarly publication repositories such as ISI Web of Knowledge, IEE Xplore, Springer, ACM, ScienceDirect, Scopus, Semantic Scholar, arXiv, etc. Based on its structure, knowledge contained in a scholarly paper is broadly classified into two major categories, which are metadata and research contributions~\cite{KEReviewAbdul2020,BrackORKGRequirements2021,jiomekongORKG2024,jiomekongLinkedOpenLitReview2024}. 

Metadata provides a brief overview of the scholarly data and contributions that contain valuable information that is beneficial to fellow researchers. AI encompassing Machine Learning (ML), Natural Language Processing (NLP), knowledge representation, and data mining provides transformative tools to automate, accelerate, and enhance the processes of scholarly knowledge search and discovery, acquisition, and organization~\cite{eger2025transformingsciencelargelanguage,bolanos2024ALiteReviews}. 

\subsubsection{Literature Search, and Discovery}
AI systems for literature search and discovery aim at accelerating the identification of relevant literature to the research topic from the vast scholarly databases, thereby reducing information overload that traditional keyword searches may cause~\cite{bolanos2024ALiteReviews, whitfield2023elicit, stephen2011library, o2014techniques, bernard2025using, heikkila2024human}. These tools go beyond conventional citations by identifying supporting, contrasting, and background references, giving a clearer picture of how a paper has been received in the scholarly community. They use NLP and Knowledge Graph (KG) analytics to provide contextualized search results, helping researchers perform comprehensive literature search, minimizing the risk of missing critical sources, and identifying knowledge gaps~\cite{bolanos2024ALiteReviews}. Key features involve:
\begin{itemize}
    \item \textbf{Semantic search:}~\cite{GuhaSemanticSearch2003, fazzinga2010semantic} traditional literature searches are often limited by keywords. Semantic search feature aims to understand the meaning and context of user queries, and retrieve relevant results even if the exact keywords are not present.
    
    \item \textbf{Papers recommendation:} the recommendation feature consists of analyzing keywords, citations, researchers' reading history, and contextual similarities to suggest related papers, ensuring that researchers have access to the most pertinent literature~\cite{bai2019scientific}. 

    \item \textbf{Citation analysis:} This feature involves analyzing existing references to suggest relevant citations, thereby helping researchers discover key studies they might have overlooked~\cite{pornprasit2022enhancing}. 

    \item \textbf{Bibliography management:} the bibliography management consists of finding and managing references, automatic organization of citations, and automatic generation of bibliographies~\cite{stephen2011library}.

    \item \textbf{Cross-disciplinary retrieval:} This involves identifying relevant literature across diverse fields by recognizing domain-specific terminologies~\cite{o2014techniques}.
    
    \item \textbf{Dynamic updating:} to be relevant, the system should continuously integrate new publications and data to maintain an up-to-date knowledge base~\cite{ImprovingAccessAuer2020}.
\end{itemize}

Several AI systems are developed in this direction: Semantic Scholar~\cite{kinney2025semanticscholaropendata}, Elicit~\cite{whitfield2023elicit, bernard2025using}, ResearchRabbit~\cite{cole2023researchrabbit}, \href{https://www.dimensions.ai/}{Dimensions AI}, \href{https://scite.ai}{Scite}, AMiner~\cite{tang2016aminer}, \href{https://www.connectedpapers.com/}{Connected Papers}.

\subsubsection{Knowledge Extraction}
Following the literature discovery, AI systems are used for (semi-)automatic identification and extraction of paper metadata and key insights consisting of authors' research contributions.

Knowledge Acquisition (KA) from scholarly data involves the extraction of structured content in the form of entities, relations, facts, terms, and other types of information that may help researchers to understand the papers and get insights from them \cite{BrackORKGRequirements2021,jiomekongORKG2024,JiomekongWWWConf2024}. It serves as a pivotal activity for extracting metadata and authors' contributions from scholarly communication. Given the vast amount of papers published yearly, the automatic extraction of scholarly data from scholarly text allows the inclusion of a number of papers related to the research topic in the study. 

Scholarly knowledge often resides in unstructured formats (text in papers, figures, tables). AI, particularly NLP and Computer Vision (CV), can extract structured data from these sources. For example, it can extract chemical properties from text, identify protein interactions from biological diagrams, or pull experimental parameters from methods sections. Key features of these systems are:
\begin{itemize}
    \item \textbf{Text classification:}~\cite{brack2022cross} consists of automatic categorization of parts of scholarly texts in order to identify and extract meaningful information.
    \item \textbf{Named entity recognition:}~\cite{d2022computer} consists of the identification and classification of scholarly entities such as authors, affiliation, methods, datasets, and tools into predefined categories.
    \item \textbf{Relation extraction:}~\cite{jiang2020targeting} consists of the identification and extraction of semantic relationships between entities mentioned in the text.    
    \item \textbf{Reading comprehension:}~\cite{baradaran2022survey} help users identify and comprehend authors contributions.
\end{itemize}
Several AI systems such as SciBERT~\cite{beltagy2019scibert}, AllenNLP~\cite{gardner2018allennlp}, scispaCy~\cite{neumann2019scispacy}, \href{https://scite.ai/}{Scite.ai}, \href{https://www.explainpaper.com}{ExplainPaper}, \href{https://chatgpt.com/}{ChatGPT}, \href{https://gemini.google.com/}{Gemini} are used for knowledge extraction.

\subsubsection{Knowledge Organization}
Acquisition is only one part of the equation. Effective organization of scholarly knowledge is crucial, as it enables its use and reuse by different stakeholders. To this end, KGs play a central role in organizing and representing scholarly knowledge by modeling entities (e.g., authors, institutions, concepts, datasets) and their semantic relationships (e.g., citations, collaborations, method usage), structuring thereby scholarly knowledge around shared vocabularies and ontologies~\cite{verma2023scholarly, ImprovingAccessAuer2020,auer2025ORKGNesy,jiomekongORKG2024, JiomekongWWWConf2024, DessiCS-KG2022, brack2020requirements, priem2022openalex,ManghiOpenAire2012,manghiOpenAire2026}. 

The growing use of KGs reflects the shift toward structured, machine-readable representations of scholarly knowledge, which are essentially for scalable, AI-driven research discovery and evaluation. Key features of AI systems for knowledge scholarly knowledge organization involves~\cite{verma2023scholarly, ImprovingAccessAuer2020,auer2025ORKGNesy,jiomekongORKG2024, JiomekongWWWConf2024, DessiCS-KG2022, brack2020requirements, priem2022openalex}:
\begin{itemize}
    \item \textbf{Structure scholarly knowledge:} entities composed of papers metadata, authors' research contributions, underlying datasets, associated code, and relevant experimental protocols are linked, forming a richer, more interconnected knowledge ecosystem.
    
    \item \textbf{Automated KGs and ontology construction:} to facilitate the extraction of entities and their relationships from millions of research papers, datasets are used to train ML models.

    \item \textbf{FAIR principle:} during scholarly knowledge organization, the FAIR (Findable, Accessible, Interoperable, and Reusable) principle should be fulfilled.
    
    \item \textbf{Navigation and visual exploration of scholarly data:} this feature consists of interactive graphs, charts, maps, or a visual interface to help understand, analyze, and navigate scholarly KGs.
    
    \item \textbf{Inference and reasoning capabilities to derive new knowledge:} the reasoning capability of KGs allows the inference of new connections or the answering of complex multi-hop questions that would be impossible with traditional databases.

    \item \textbf{Identifying gaps and conflicts in knowledge:} This feature involves analyzing the structure and content of scholarly KG to identify areas where knowledge is missing, contradictory, or where new connections can be inferred.
\end{itemize}

Numerous scholarly knowledge graphs are developed and used: the Open Research Knowledge Graph (ORKG)~\cite{ImprovingAccessAuer2020}, OpenAlex~\cite{priem2022openalex}, AMiner~\cite{tang2016aminer}, CovidPubGraph~\cite{pestryakova2022covidpubgraph}, Microsoft Academic Graph (MAG)~\cite{WangMicrosoftGraph2020}, Semantic Scholar KG~\cite{Wade2022TheSemanticScholar}. For instance, the OpenAIRE Graph is a large-scale open scholarly knowledge graph that aggregates and interlinks publications, datasets, software, projects, organisations, funders, and researchers from thousands of sources. Enriched through AI and full-text mining, it increasingly serves as a trusted foundation for AI agents, enabling transparent, provenance-aware scholarly discovery and research intelligence~\cite{ManghiOpenAire2012,manghiOpenAire2026}.

% Thank you so much!

\subsubsection{Community Engagement}
Realizing the full potential of AI requires not only technical innovation, but also the active involvement of researchers and scholarly communities to ensure accuracy, inclusivity, and ethical stewardship. Therefore, it is important to keep humans involved in describing articles and validating knowledge (Human-in-the-loop)~\cite{Enqvist2023Human,Walter2023Human}.

To facilitate community engagement during scholarly knowledge extraction and organization, the ORKG observatory has been developed. It allows a whole community to collaboratively build the state-of-the-art in a research domain~\cite{jiomekongORKG2024}. To this end, scholarly data such as scientific papers, books, and datasets are collected, compared, and summarized.

\subsection{AI systems for Knowledge Generation and Editing}
Present AI systems are used in the full scholarly communication workflow for generating scholarly knowledge~\cite{eger2025transformingsciencelargelanguage,bolanos2024ALiteReviews,van2023artificial}. During the ideation activity, new ideas are generated, developed, and communicated. From ideas, testable statements that propose a possible explanation for a phenomenon or predict the outcome of an experiment can be stated as a research hypothesis. One or several research questions that define what the researcher wants to investigate may emerge. This often forms the foundation for hypotheses or research objectives. These elements are used to shape the research design, methodologies, methods, tools, and experiments to conduct. The methodology explains which methods are used, why they are chosen, and how they help to answer the research question or test the hypothesis. During the experiment, implementation, or simulation, the method or methodology described is used to translate theoretical concepts into practical steps that can be observed, measured, and evaluated. Models are used to replicate real-world processes or phenomena in a controlled virtual environment. This allows researchers to study complex situations that are difficult, expensive, dangerous, or impossible to observe directly and reduces the need for physical prototypes or large-scale trials. Research results present findings or outcomes that emerge as a result of the experiment, implementation, simulation, or data analysis. The analysis of research results is used to connect the raw results to the broader context of the study. During this activity, the comparison with the previous study allows the identification of similarities, differences, and potential reasons for the results obtained. These elements, together with related work drawn from papers from which knowledge was extracted and organized, form the foundation of the research paper. During these activities, several AI systems may be used for knowledge generation and editing. The key features of these AI systems are:
\begin{itemize}
    \item \textbf{Brainstorm and explore a range of possible directions, questions, ideas, and hypotheses}. This feature consists of exploring millions of scholarly resources to answer research questions, generate novel hypotheses by analyzing complex relationships within data, and identify gaps in current knowledge, emerging areas, and under-explored areas.

    \item \textbf{Drafting:} This feature aims to speed up manuscript preparation by generating initial drafts of papers, books, master's theses, initial responses to reviewer comments, and presentations.
    
    \item \textbf{Content editing:} content editing feature aims at assisting authors during the improvement of the writing quality and clarity by paraphrasing, improving grammar, style, and coherence, and reference insertion.
    
    \item \textbf{Summarization:} This feature involves processing thousands or millions of scholarly data, including books, scientific papers, patents, and technical reports, to generate concise summaries, extract key findings, methodologies, and relationships. This enables a quick grasp of complex research without needing to read every detail, drastically reducing the time spent on initial literature surveys.
    
     \item \textbf{Tables, pictures, charts generation:} This feature involves creating data visualizations, diagrams, or synthetic images to illustrate concepts based on textual descriptions or numerical data.

     Data and results analysis, appropriate comparison tables, pictures, and charts can be generated from numeric data or a textual description of the desired visualization. The best chart type can also be suggested (e.g., bar, line, scatter, heatmap) based on data characteristics and the message to convey. Diagrams or molecular structures can also be generated.

     This feature is beneficial because it reduces manual effort in creating visuals, ensures data accuracy in figures, helps non-designers create professional-looking charts, and allows for rapid integration of visual concepts.

     \item \textbf{Image and video generation and enhancement:} This feature involves enhancing the quality of existing images (e.g., research photos, microscopy images) or creating entirely new illustrative visual assets based on textual prompts. This includes tasks like upscaling resolution, de-noising, enhancing contrast, or even creating synthetic illustrative images or short animations.

     \item \textbf{Handling big data:} research is generating petabytes of data. This feature involves analyzing the data to identify subtle correlations, anomalies, and trends that would be invisible to traditional methods. This aims at accelerating the data analysis phase.

     \item \textbf{Data analysis and pattern recognition:} this feature consists of automatic analysis of scholarly data and pattern recognition. Numerous models show powerful performance on tasks such as image recognition, analysis of complex chemical spectra, and identification of behaviors in large video datasets.

     \item \textbf{Automatic construction of ontologies and knowledge graphs:} Given the vast amount of scholarly knowledge available, manual construction of ontologies and KGs is not an easy task. This feature consists of automatic identification, extraction, and organization of entities (authors, publications, research contributions, etc.), as well as the relationships between them in a scholarly knowledge graph.

     \item \textbf{Source code generation:} this feature aims at assisting the academic community in coding, debugging, or optimizing source code used for analyses or simulation within their papers.

     \item \textbf{Content translation and adaptation:} This feature aims at reducing language barriers, particularly benefiting non-native English speakers. It supports multilingual communication.

    \item \textbf{Validation and reproducibility:} This feature consists of verifying the correctness of statistical analyses and computational workflows. It allows for enhanced transparency and rigor by automating checks for methodological soundness and enabling the sharing of reproducible research components.

    \item \textbf{Chatbots and virtual assistants:} This feature aims to help authors interact with the AI system, providing instant answers based on existing studies.
    
    \item \textbf{Scientific papers generation:} this feature consists of the automatic generation of the whole research papers.
\end{itemize}

AI systems for knowledge generation and editing can be classified into: general AI tools such as ChatGPT, Gemini, DeepSeek, and Claude AI, which cover almost all the features presented above. Specialized tools that involve only several features. For instance, \href{https://www.grammarly.com/}{Grammarly}, \href{https://www.trinka.ai/}{Trinka AI} are specialized on content editing; \href{https://www.microsoft.com/en-us/power-platform/products/power-bi}{Power BI}, \href{https://plotly.com/}{plotly}, \href{https://lookerstudio.google.com/}{Google Data Studio}; \href{https://www.adobe.com/uk/sensei/generative-ai.html}{Adobe Sensei}, \href{https://openart.ai}{DALL-E}, \href{https://www.topazlabs.com/}{Topaz Labs}, \href{https://www.imagine.art/}{Midjourney}, \href{https://www.synthesia.io}{Synthesia} for images and video generation and enhancement; \href{https://deepmind.google/science/alphafold/}{Alphafold} for handling big data; \href{https://www.statreviewer.com/}{StatReviewer}, \href{https://codeocean.com/}{Code Ocean} for validation and reproducibility; \href{https://sakana.ai/}{Sakana AI}, TIB assistant.

\subsection{AI systems for Peer Review, Publication, and Post Publication}
Peer review happens in two phases: (1) Firstly, scholarly resources are assigned to reviewers, and then, the reviewers review and submit their comments. Given the number of scholarly resources submitted for publication in some conferences (11,239 papers were submitted to ICCV 2025) and the variety of research topics, there is a need for tools that will facilitate the paper attribution to reviewers, avoiding conflicts of interest and ensuring that papers are assigned to the appropriate reviewer. These systems are used by editorial boards for the following purposes: 

\begin{itemize}
    \item Assess submissions for reproducibility
    \item Ethics compliance
    \item Plagiarism
    \item Methodological flaws
    \item Reviewer recommendation
    \item Bias detection in review.
\end{itemize}

Several AI tools have been developed in this direction: \href{https://www.penelope.ai/}{Penelope.ai}, \href{https://www.ithenticate.com/}{iThenticate}. (2) Secondly, reviewers assess the relevance of the research, methodology, data, experiments, results, and conclusion of the scholarly work. To this end, several AI tools can be used for knowledge generation and editing, as presented above. In addition, several tools such as \href{https://sciscore.com/}{SciScore} can assist the reviewer during the review writing.

After the publication activity, AI systems may be used to scale up the dissemination, metadata enrichment, and track the broader impact of the research. These systems should be able to:
\begin{itemize}
    \item Broadening the reach and improving the accessibility of scholarly communication to diverse audiences, beyond just academic peers.
    
    \item Automated translation: translating research papers into multiple languages to break down linguistic barriers and facilitate global knowledge sharing.
    
    \item Plain language summaries (PLS): automatically generating simplified, accessible summaries of complex scholarly resources for policymakers, journalists, educators, or the general public. 
    
    \item Adaptive learning: creating personalized educational content from research findings, tailored to a learner's background and progress.
    
    \item Multi-format content generation: automatically adapting research content into different formats (e.g., audio summaries, concise social media posts, interactive QA interfaces) for broader dissemination.    
\end{itemize}
Examples of tools currently used are: Altmetric~\cite{trueger2015altmetric} and PlumX~\cite{champieux2015plumx}.

\subsection{AI systems for Slide Presentation}
After meticulous analysis, scholarly content preparation, and then publication, effective communication of the content to diverse audiences is crucial for impact.  This is generally done in conferences, hackathons, etc. AI is increasingly offering powerful tools to enhance this activity, moving beyond simple content generation to sophisticated visual, auditory, and interactive presentation aids in conferences. Key features of existing AI used include:
\begin{itemize}
    \item \textbf{Content generation:} This feature involves creating or refining the core textual and visual components of a presentation. Several types of content, such as slides, posters, voices, and subtitles, can be generated to save time.

    \item \textbf{Content refinement:} This involves grammar correction, optimization of bullet points, suggestions for impactful phrasing, and ensuring consistency in terminology for specific audiences, such as non-experts.
    
    \item \textbf{Layout design:} This feature assists in the aesthetic and structural organization of the presentation. It suggests or applies aesthetically pleasing layouts, font combinations, colour schemes, and visual hierarchies for slides based on content, brand guidelines, or desired mood. It allows you to transform raw notes into visually coherent slides.

    \item \textbf{Visual storytelling:} this feature consists of the analysis of the presented content (text, data, images). It suggests optimal ways to break it down across slides, arrange elements for maximum impact, grouping related bullets points, recommend moving large blocks to supplementary slides, indicate where an image would enhance a specific point and create a coherent visual narrative.

    \item \textbf{Speech coaching and delivery feedback:} consists of the analysis of the presenter's voice (pitch, pace, volume), body language (via webcam), and verbal content during practice sessions to provide real-time or post-session feedback. It helps presenter refine their delivery, build confidence, and ensure clarity and engagement for the audience, leading to an impactful talk. 
    
    \item \textbf{Real-time captioning, translation, and accessibility:} This feature provides real-time closed captions for spoken content during live presentations, instant translation of spoken or textual content into multiple languages, and the generation of sign language avatars.

    \item \textbf{Audience engagement analytics:} for visual presentation, this feature analyzes audience participation metrics (e.g., chat activity, poll responses, virtual hand raises, even gaze direction if consent is given) to provide presenters with insights into engagement levels, allowing him to adapt their delivery or content for future presentations and assess impact.

    \item \textbf{Interactive data visualization and exploration:} this feature allows presenters to create interactive visualizations where audience members can drill down into data, filter results, or change parameters in real-time, moving beyond static charts
\end{itemize}
Generative AI  tools such as ChatGPT, Gemini, DeepSeek, and Claude AI cover several features presented above while some tools are specialized for several features. For instance, \href{https://www.beautiful.ai/}{Beautiful.ai}, Powerpresent AI\footnote{\url{https://www.powerpresent.ai/}} for slide generation and refinement; \href{https://support.microsoft.com/en-us/office/rehearse-your-slide-show-with-speaker-coach-cd7fc941-5c3b-498c-a225-83ef3f64f07b}{PowerPoint presenter coach} for speech coaching and delivery feedback; \href{https://hopin.com/}{Hopin} and \href{https://remo.co/}{Remo} for audience engagement analytics.

\section{Using Scholarly Knowledge and Impact}
Scholarly communication's utility spans various domains, from deepening our fundamental comprehension to driving practical solutions for real-world problems. These solutions often arise through translational research, which converts basic scientific findings into usable technologies, or through partnerships between academic institutions and industry, which are essential for advancing the SDGs~\cite {UNSDGs2015}. Achieving the United Nations (UN) Sustainable Development Goals (SDGs) ambitions hinges critically on the effective generation, dissemination, and utilization of scholarly knowledge.

The relationship between the SDGs and scholarly communication is symbiotic. On one hand, scholarly communication is the engine that translates academic work into actionable knowledge, making it a critical enabler for progress on the SDGs. On the other hand, SDGs are profoundly shaping the landscape and priorities of scholarly communication. During this symbiotic relationship, AI offers unprecedented opportunities to real-time understanding and implementation of insights. To ensure that AI enhances rather than distorts the use of scholarly knowledge, it is essential to establish a robust framework for evaluating AI-driven research.

\subsection{Use of Scholarly Knowledge and Impact}

\subsubsection{Use of Scholarly Knowledge in Education}
SDG 4 ensures inclusive and equitable quality education and promotes lifelong learning opportunities for all. It is driven by scholarly knowledge and has a direct impact on all the other SDGs. The rapid development of AI systems has significantly transformed the educational field. This increasing trend was accelerated during the COVID-19 pandemic, when remote learning became widespread and reliance on digital technologies reached unprecedented levels~\cite{han2022analysis, vizconde2024generative}. As a result, AI systems have become integral components of educational systems, utilized by administrators, teachers, and students. The widespread availability and increasing accessibility of these systems have created more equitable access to knowledge globally. This spread is particularly significant for countries and regions that lack access to top-tier educational resources, helping to bridge the educational gap between developed and developing nations~\cite{alsagri2024evaluating}. On one hand, KGs such as the ORKG are used to organize scholarly data to provide teachers and students with synthetic views. Shared using the FAIR principle, this ensures transparency, reproducibility, and enables further analysis by students and teachers.

On the other hand, Generative AI such as ChatGPT, Gemini, DeepSeek, and Claude AI are used to synthesize complex academic work into accessible formats that are tailored for students and teachers. For instance, a set of academic resources (books, research papers, etc.) in English can be uploaded into an AI system, which then provides answers on objective, methodology, results, etc., using the language of the user. AI can also be used to generate plain language summaries, create explanatory visualizations, and provide translation in the user's language. For learners with disability, AI systems can offer support through voice or video for scholarly data illustration~\cite {vizconde2024generative, kaur2021qualitative}.

%%%%%%%%%%%%%%%%%%%%%%%%%%%%%%%%%%%%%
\subsubsection{Using Scholarly Knowledge in Food science and Nutrition}
According to WHO, every country in the world is affected by one or more forms of malnutrition~\cite{malnutritionWHO2023}. Understanding Food information enables people to maintain a healthy diet, enhance productivity and crop quality, and develop systems for food search, recommendation, and question answering. To this end, AI systems are employed through the entire process of food information engineering - from acquisition and processing to the dissemination of up-to-date food information to various stakeholders~\cite{JiomekongFoodInfoEnginee2023,JiomekongFoodInfoEnginee2023}. Given the large number and diverse type of scholarly resources containing food data, several AI systems are used for automatic extraction and processing~\cite{collectFoodInfoJiomekong2023,ReviewFoodInforProcess_Azanzi_2023}. Once acquired, food ontologies such as FoodOn~\cite{dooley2018foodon}, food knowledge graphs such as FoodKG~\cite{haussmann2019foodkg} and food linked data such as FOODpedia~\cite{kolchin2015foodpedia} - AGROVOC~\cite{caracciolo2013agrovoc} are used to provide a shared understanding of food knowledge or tackle data harmonisation problems.

ML models such as neural networks (NNs), are used to learn associations from food data and build models such as food recognition (e.g., VGG-19, AlexNet, GoogLeNet, Resnet-50, DenseNet, MobileNets, ShuffleNets, etc.)~\cite{FRecoDiaPatAnthimopoulos2014, bettadapura2015leveraging}, food recommendation~\cite{stefanidis2022protein, Brintha2022AFR, chen2021personalized}, ingredient substitution~\cite{kim2024survey}, automatic fruit classification and counting~\cite{zhang2022complete} as well as plant disease detection in large farms~\cite{hassena2023quality, liu2021plant, bakr2022densenet}. AI/ML models have also been developed and deployed to increase agricultural productivity while reducing adverse impacts on the environment~\cite{visentin2023mixed, alazzai2024precision}, and answer food QA problems~\cite{chen2021personalized}.

%%%%%%%%%%%%%%%%%%%%%%%%%%%%%%%%%%%%%%%%%%%
\subsubsection{In Physics}
Artificial intelligence is revolutionizing physics-based engineering applications across multiple critical domains. The field employs AI technologies ranging from traditional machine learning approaches to cutting-edge deep learning architectures. In structural engineering and seismic control, AI technologies including reinforcement learning, deep neural networks, and genetic algorithms~\cite{fu2020fuzzy,rahmani2019framework,greco2017seismic} are enabling model-free intelligent control systems that adapt to earthquake uncertainties without requiring precise system models. Smart materials and vibration control benefit from fuzzy-neural network controllers that overcome the limitations of conventional systems by accurately modeling nonlinear relationships in magneto-rheological elastomer systems~\cite{fu2020fuzzy}. Wind engineering applications leverage machine learning models including Light Gradient Boosting Machines, CNNs, and Generative Adversarial Networks to predict wind loads, crosswind effects, and flutter performance in tall buildings and bridges~\cite{deng2020concrete,lin2022machine}, significantly reducing the need for costly wind tunnel testing.

The integration of AI system in the physic domain have significant positive impacts across multiple SDGs~\cite{addagada2022role}: SDG 9 (Industry, Innovation and Infrastructure) is directly advanced through AI-enhanced structural design, automated construction quality control, and intelligent infrastructure monitoring systems. AI enables the development of more robust infrastructure capable of withstanding natural disasters and adapting to changing environmental conditions; SDG 11 (Sustainable Cities and Communities) benefits from AI applications in urban infrastructure management, including smart building systems that optimize energy consumption, automated monitoring of infrastructures like bridges and tunnels; SDG 13 (Climate Action) is supported through AI-optimized structural designs that minimize material usage and environmental impact, intelligent systems that improve building energy efficiency, and enhanced disaster preparedness capabilities that help communities adapt to climate change impacts.

Despite significant advances, several challenges persist in applications for Physics. Data quality and availability remain critical issues, as many AI models require large, high-quality datasets that are expensive and time-consuming to collect in engineering contexts.

%%%%%%%%%%%%%%%%%%%%%%%%%%%%%%%%%%%%%%%%%%%
\subsubsection{In Environment}
The vast corpus of scholarly literature available today, combined with AI's ability to extract and synthesize information from thousands of research papers has enhanced environmental decision-making and policy development~\cite{konya2024recent}. The explainability capacity of Natural Language Processing (NLP) systems has become important in environmental decision making. These tools can process and analyze extensive scholarly databases, tracking environmental data trends and making complex research findings accessible to policymakers and stakeholders, thereby bridging the gap between academic research and practical environmental action.

The increasing global availability of datasets and scholarly research on clean water and sanitation, as well as affordable and clean energy, has created unprecedented opportunities for AI driven environmental solutions. Big data analysis powered by AI systems such as Claude and GPT enables the processing of massive environmental datasets, providing policymakers with timely and accurate information for informed decision making~\cite{konya2024recent}. This capability directly supports the achievement of: SDG 6 (Clean Water and Sanitation): AI systems can analyze water quality data, predict contamination patterns, and optimize water treatment processes~\cite{durgun2024real,singh2024prediction}; SDG 7 (Affordable and Clean Energy): Machine learning algorithms can optimize renewable energy distribution~\cite{saxena2024hybrid}, predict energy demand~\cite{bai2024news}, and improve grid efficiency~\cite{yu2022transfer}.

%%%%%%%%%%%%%%%%%%%%%%%%%%%%%%%%%%%%%%%%%%%
\subsubsection{In Economics}
AI offers substantial productivity improvements for economists by automating micro-tasks and routine processes. Key applications include research classification, economic forecasting and indicator prediction, data analysis and processing, ideation and feedback for research, background research automation, coding assistance, and mathematical derivations~\cite{korinek2023generative,dong2024scoping}. The field primarily utilizes Large Language Models (LLMs) such as ChatGPT, GPT-3.5, GPT-4, Microsoft Copilot, Google's Gemini and Bard, PaLM-2, Claude 2, and LlaMA 2~\cite{korinek2023generative}. In business related domains like accounting and finance, AI is being used for classification tasks, sentiment analysis, text summarization, and content generation~\cite{fisher2016natural}. For economic forecasting specifically, various machine learning techniques are employed including Artificial Neural Networks, adaptive neuro-fuzzy inference systems, genetic programming, support vector regression, and extreme learning machines~\cite{pakholchuk2024methodological}. 

The application of AI in economics has significant implications for achieving multiple Sustainable Development Goals (SDGs). Most directly, AI's role in enhancing economic forecasting and strategic planning supports SDG 8: Decent Work and Economic Growth by enabling more informed policy decisions and efficient resource allocation. The automation of routine economic tasks and improved productivity can contribute to sustained economic growth while potentially creating new types of employment opportunities~\cite{bali2020achieving}.
AI can also support SDG 10: Reduced Inequalities by making sophisticated economic analysis accessible to smaller institutions and developing economies that previously lacked such capabilities~\cite{bali2020achieving}.

%%%%%%%%%%%%%%%%%%%%%%%%%%%%%%%%%%%%%%%%%%%
\subsubsection{In Law}
AI, particularly Large Language Models (LLMs) like GPT, BERT, LEGAL-BERT is being applied across numerous legal tasks including legal judgment prediction, document analysis and drafting, contract review, case prediction, regulatory compliance, legal research, and even providing legal advice~\cite{siino2025exploring}. These technologies are also used for document summarization, information extraction, and automated decision-making processes in the field of Law which as other fields, is not exempt of the voluminous amount of data (legal text, proceedings, etc) available. The availability of a voluminous dataset such as: ECHR (European Court of Human Rights, ILDC (Indian Legal Documents Corpus), UNFAIR-ToS (Unfair Terms of Service) is also a plus since it provides enough material for the training of AI models~\cite{sun2023short}.

The integration of AI in law offers several advantages. It can automate routine legal tasks, making legal services more efficient and potentially more affordable and accessible~\cite{magnusson2019legal}. AI systems demonstrate high accuracy in statutory reasoning and can effectively manage the complexity of legal language~\cite{katz2023natural}. They also show promise in enhancing legal practice through improved document processing and analysis capabilities.

The application of AI in the legal field has significant impact onc85 SDGs, particularly SDG 16: Peace, Justice and Strong Institutions. By making legal services more accessible and affordable through automation, AI directly contributes to ensuring equal access to justice for all, which is a core target of SDG 16. The technology's ability to process vast amounts of legal information efficiently can strengthen institutional capacity and improve the rule of law. Additionally, AI in law supports SDG 10: Reduced Inequalities by potentially broadening access to legal knowledge and services that are mainly available only to those who can afford expensive legal counsel. The automation of routine legal tasks could bridge the justice gap, particularly benefiting marginalized communities who face barriers to legal representation~\cite{meskic2022digitalization}.

However, the bias and fairness concerns identified in AI systems pose risks to these same SDGs. If not properly addressed, AI could perpetuate or even amplify existing inequalities in the justice system, diminishing efforts to achieve fair and inclusive institutions. The privacy and transparency challenges also relate to SDG 16's emphasis on developing effective, accountable, and transparent institutions at all levels~\cite{meskic2022digitalization}. Despite the potential, several major obstacles persist. The inherent ``open texture" of law - characterized by vague and ambiguous rules makes automation particularly challenging~\cite{katz2023natural}. AI systems still struggle with the complexity and variability of legal language, and there's a persistent need for large, high-quality annotated datasets~\cite{magnusson2019legal}. The ``black box" nature of many AI systems also creates difficulties in explaining decision-making processes, which is crucial in legal contexts~\cite{sun2023short}.

%%%%%%%%%%%%%%%%%%%%%%%%%%%%%%%%%%%%%%%%%%
\subsubsection{Improving Collaboration and Communication}
SDGs are inherently interdisciplinary. Thus, achieving SDGs demands interdisciplinary, cross-border cooperation that is grounded in scholarly evidence.  This cooperation will have a direct impact on SDG 17 (Partnership for the Goals). AI may be used to build scholarly communication networks involving researchers and  conferences, thus, connecting researchers across geographical and institutional boundaries and enabling shared efforts to address global challenges. To make research results accessible to different stakeholders, AI may be used personalize content delivery, automate translation, and simplify complex scholarly language into more accessible terms for policymakers, journalists, educators, and the general public. For instance, AI can rapidly gather, filter, and synthesize evidence from multiple studies to  answer specific policy questions (e.g., ``which foods is appropriate to children with vitamin A deficiency?"). This provides policymakers with comprehensive, evidence-based briefings. AI-powered translation tools may be used to facilitate real-time collaboration between researchers from different linguistic backgrounds.

%%%%%%%%%%%%%%%%%%%%%%%Evaluation
\subsection{Evaluation of the AI-driven use of Research Results}
The emergence of AI-driven research necessitates the development of new interdisciplinary methodologies for evaluation that blend traditional scholarly validation with AI-specific assessment techniques.

\subsubsection{Evaluation of AI Models}
The models used during knowledge discovery, knowledge extraction and generation should be evaluated using context-specific metrics. At the technical level, accuracy, precision, recall, F1 score, MEAN, ROUGE, BLEU, AUC measures should be used. At the operational level, speed, scalability, cost-effectiveness are applicable. At the social level, fairness, accountability, transparency, and ethical acceptability should be applied.

The ``black box" nature of many advanced AI models necessitates a deeper look into their internal workings and characteristics. To allow models to be easily evaluated, they should be reproducible and transparent.  They should involve open datasets and codebases, with clear documentation of model architectures and parameters, reporting bias and model limitations.   These systems should fulfill the following aims:

\begin{itemize}
    \item Interpretability~\cite{Ru2021InterpretableNA, mills2024autobuild} and explainability (XAI)~\cite{mersha2024explainable} aims at providing human-understandable explanations for AI decisions, revealing which data features influenced a particular outcome. For instance, which documents were used to generate a content. 
    \item Robustness aims to evaluate the capacity of the model when the data is slightly different or noisy. 
    \item Fairness and bias aims to check for algorithmic bias, concerning demographic groups. In the scholarly communication, the question will be to know if the AI system consider the scholarly data from different geographical region in the world. 
    \item Uncertainty quantification  provide the level of confidence in an AI system's predictions. However, currently, AI systems used in scholarly communication do not provide this information.  This is problematic because these systems hallucinate at times. 
    \item Computational efficiency and resources aims at assuring that the need in energy consumption, specialized hardware do not limit its broader adoption or sustainability. 
    \item Reproducibility and replicability aims for an AI model to be documented, and ideally made available so as to allow fellow researchers to reproduce the AI's training process and obtain similar results from the same data. 
    \item Validity aims to measure if the AI-generated conclusions, prediction, ... truly accurate and reflective of reality. This means rigorously testing the AI's outputs against human-verified ``ground truth" data or established benchmark dataset. 
    \item Reliability measures the performance consistent over and across different, yet relevant, inputs, or slightly perturbed data. A reliable model should not yield significant different results for minor, imperceptible changes in input data.
\end{itemize}

\subsubsection{Evaluation of Scholarly Communication Supports}
Given that AI systems may be used to generate scholarly communication support artifacts such as books, reports, scientific papers, datasets, images, presentation, and graphs, the evaluation of those supports should be considered. The traditional peer-review process, the long-standing cornerstone of scholarly quality control and validation, must evolve significantly to accommodate AI-driven research. New guidelines for reviewing AI-driven research should emerge and peer review should consider AI expertise. Reviewers must possess not only deep domain expertise in the scholarly domain, but also a robust and critical understanding of the use of AI for knowledge extraction, generation and editing to critically assess the appropriateness of the AI generated content, and the contributions of the authors, etc. They should be able to determine what should be accepted for the conference/journal and  how to detect the use of AI, etc. This dual expertise is crucial to rigorously assess the ethical considerations and responsible use of AI tools throughout the research.

Journals are increasingly developing and publishing specific guidelines for reviewing AI-driven research while emphasizing AI methodology transparency, ethical considerations, and documented impact. This may necessitate a shift towards more interdisciplinary review panels.

\subsubsection{Impact Assessment Based on SDGs}
Given the large volume of scholarly communication published yearly, the question on the evaluation of the impact of these data is posed. One of the best ways to evaluate will be to assess the impact of research on SDGs. The ORKG, for instance allows researchers to assign which SDGs a scientific paper impacts. Currently, this task is still manual and many papers indexed by the ORKG do not provide this information. However, AI can help analyze the real-world impact of implemented policies or interventions, evaluate the prediction of future trends or risks such as the spread of infectious diseases, the impact of climate change on specific regions, or the economic effects of technological shifts. 

\subsubsection{Ethical Evaluation}
Beyond technical performance, the broader societal implications of the integration of AI in the research life cycle must be evaluated. The societal impact of the research output will measure the potential for the misuse of the research results. The accountability aims to determine who is accountable for errors or harmful outcomes when AI contributes significantly to research findings. 

The power of AI to synthesize vast amounts of data and identify novel patterns raises fundamental questions about its ethical deployment. AI models learn from the data they are trained on. If this data reflects historical, societal, or sampling biases, the AI will not only perpetuate these biases but can also amplify them in its generated knowledge. For instance, an AI trained predominantly on one demographic group might generate research insights that are less relevant or even harmful to others. If these biases go unchecked, AI can reinforce and legitimate systemic injustices. Addressing this requires inclusive datasets, bias audits and impact assessment.

\section{Ethics and Responsibility}
\begin{figure}
    \centering
    \includegraphics[scale=0.3]{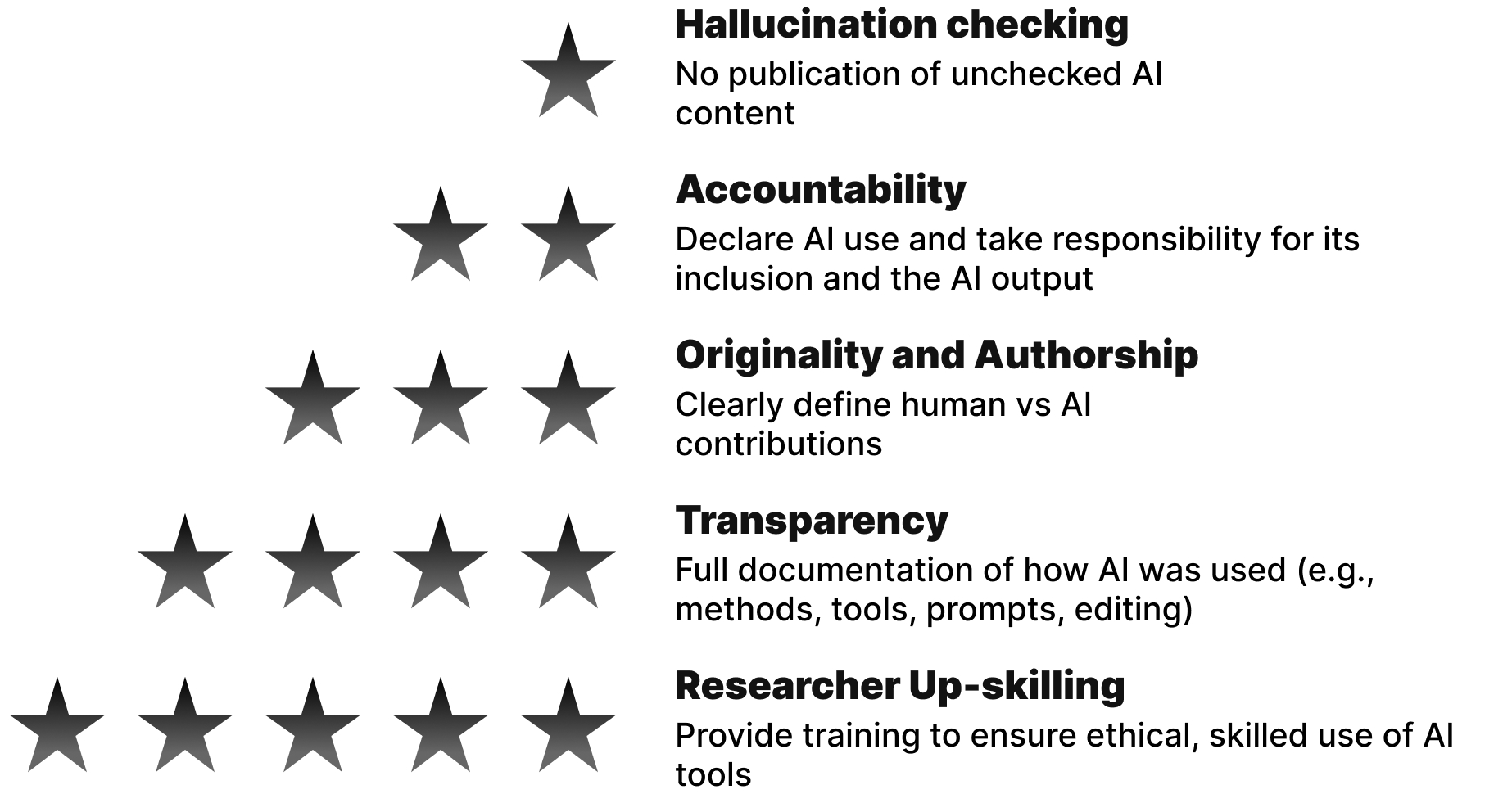}
    \caption{Illustration of the five stars ethics on AI for scholarly communication}
    \label{fig:fiveStartEthics}
\end{figure}

As AI becomes a sophisticated tool for knowledge extraction, knowledge generation, content creation, autonomous discovery, etc. critical questions arise regarding reliability, transparency, fairness, accountability, originality, authorship. Responsible use of AI in research demands deliberate ethical reflection, institutional guidelines, and continuous oversight that extend beyond traditional research ethics. Following the five-star model proposed by Tim Berners-Lee for open data, we designed the framework presented in Fig. \ref{fig:fiveStartEthics} to illustrate how five-start AI ethics in scholarly communication can be structured.

%%%%%%%%%%%%%%%%%%%%%%%%%%%%%%%%%%%%%%%%%%
\subsection{Transparency}
Transparency is a foundational ethical principle in scientific research and a prerequisite for reproducibility, one of the core pillars of the scientific method. As generative AI systems become increasingly integrated into various stages of the research lifecycle, from data analysis to content generation, it is imperative that their use be clearly disclosed. This includes providing detailed information on prompts, tool versions, configurations, and any other relevant parameters that could influence the outputs.

Failure to disclose the role of generative AI, or the minimization of its contributions, risks undermining the integrity and reproducibility of the research. Full transparency enables peer reviewers and readers to assess the provenance of research findings and ensures accountability in the use of automated systems.

Current editorial guidelines in many journals recommend that authors explicitly declare the use of generative AI tools when such tools have made substantial contributions to the research process, such as generating content, conducting data analysis, or assisting in the interpretation of results. In contrast, minor uses—such as grammar correction or rephrasing—typically do not require formal disclosure. Nonetheless, as the capabilities of AI tools evolve, clearer and more standardized guidelines may be needed to ensure consistent and ethically sound reporting practices.

At a minimum, transparency in scientific research requires that authors provide a clear and detailed description of their methodology. However, in practice, this requirement is often insufficient—particularly for complex AI projects involving dozens of dependencies, hundreds of files, and thousands of lines of code. Capturing the full intricacy of such systems within the confines of a typical ten-page, double-columned scientific manuscript presents a significant challenge.

To address this limitation, leading AI venues such as AAAI and NeurIPS have introduced reproducibility checklists~\cite{neurips2021Checklist} that prompt authors to report additional details relevant to transparency, such as dataset availability, hyperparameters, code structure, and compute resources. Many conferences also encourage or require the inclusion of source code, which can aid reviewers in assessing the validity and robustness of submitted work. Despite these efforts, reproducibility artifacts typically remain part of the submission’s supplementary materials, and their evaluation is often left to the discretion of reviewers~\cite{neurips2025C4P}. Given the growing volume of submissions to top-tier AI conferences~\cite{xin2025conf}, e.g., over 25k submissions for NeurIPS'25~\cite{CSPaper_2025}, it is not guaranteed that these supplementary components will meaningfully influence the review process or acceptance outcomes.

In contrast, organizations such as the Association for Computational Machinery (ACM)~\cite{acm2021ReviewBadges} and USENIX take matters a step further by explicitly denoting accepted manuscripts that pass specific transparency and reproducibility thresholds with high-visibly badges on the on the top-right corner of a manuscript's first page in order to foster attention and boost impact. Examples of this process include ANSOR~\cite{zheng2020ansor} and Optimus~\cite{cai2021optimus}. Such assurances provide a greater degree of transparency, for while the peer-review process primarily confirms the novelty/merits of a work, post-acceptance evaluation offers a chance to further confirm implementation correctness and incentivize authors to be more transparent and rigorous in their practices. 

\subsection{Accountability}
AI systems used to assist researchers in scholarly communication often operate as ``black boxes,” offering little visibility into how decisions are made or outputs are produced. This opacity raises critical questions of accountability. For example, if an AI system autonomously designs experiments, analyzes data, proposes novel theories, or generates hypotheses, who is held responsible if the resulting research is flawed, unethical, or harmful? Similarly, if such a system was trained on copyrighted material without proper attribution, who is liable for any derived content it generates? Ultimately, human researchers, their affiliated institutions, and funding bodies must retain full accountability—not only for their decisions but also for the tools they employ, including AI-generated content. They are responsible for addressing ethical concerns, ensuring the accuracy and fairness of AI-generated contributions, complying with data protection and privacy regulations, and observing copyright and intellectual property norms.

These responsibilities necessitate the development and adoption of clear frameworks for human oversight, continuous review, and transparent reporting in AI-assisted research. Even in highly automated workflows, human-in-the-loop decision-making, validation, and ethical checkpoints remain indispensable. This principle requires that researchers actively interrogate, critique, and verify the outputs of AI tools, rather than passively accepting them.

Efforts such as Verifiable EL Reasoner~\cite{ileri2024vel} exemplify how formal verification methods can contribute to this goal. By ensuring that logical inferences made by AI systems are both sound and transparent, such tools provide a foundation for accountable and explainable AI in research contexts. They help bridge the gap between powerful automated reasoning capabilities and the ethical imperatives of human oversight. Current publication practices in leading conferences and journals reinforce this view by placing ultimate responsibility for AI-generated content on the authors. However, as AI becomes increasingly integrated into the research pipeline, the need for verifiable, transparent, and ethically governed AI tools will only grow.

%%%%%%%%%%%%%%%%%%%%%%%%%%%%%%%%%%%%%%%%%%
\subsection{Originality, Plagiarism, and Authorship}
Upholding the integrity of the scientific record requires sustained human vigilance and a deep commitment to quality, transparency, and accountability. Regardless of whether AI systems are used in the research process, it is ultimately the responsibility of human researchers to ensure that all scholarly content they produce is original, properly cited, and intellectually honest. This includes manually verifying the accuracy, appropriateness, and originality of AI-generated content.

As generative AI becomes more capable of producing fluent and coherent text, efficiently synthesizing visual content~\cite{zheng2024open, mills2025qua2sedimo, feng2025q, chen2025fp4dit}, solving challenging reasoning problems~\cite{liu2024deepseek, ghasemabadi2025guided} and even full research drafts~\cite{gottweis2025towards, lu2024ai}, the peer-review process must evolve accordingly. Reviewers and editors must look beyond surface-level linguistic fluency and assess the conceptual soundness, methodological rigor, and genuine novelty of submissions. The potential for AI-generated content to mimic academic style without contributing substantive insight raises serious concerns about research integrity.

When substantial portions of a manuscript are generated by AI without proper citation or disclosure, it may constitute a form of plagiarism or academic misconduct. This is particularly important given that generative AI systems do not create content from first principles; rather, they synthesize outputs based on patterns learned from existing texts, which may include copyrighted or proprietary materials.

The rise of generative AI poses profound ethical questions for scientific communication and authorship. While many leading publishers—including Elsevier, Nature, and Science—have issued guidelines prohibiting the listing of AI systems as authors on the grounds that they cannot assume responsibility for the work, debates continue around the role of AI in the authorship process. Some scholars advocate for the acknowledgment of ``AI-assisted authorship” when models contribute significantly to content creation, so long as their involvement is transparently disclosed and human accountability is clearly maintained.

In this rapidly evolving landscape, maintaining the credibility of scientific publishing demands clear ethical frameworks, informed editorial oversight, and a collective commitment to preserving the authenticity of the scholarly record.

%%%%%%%%%%%%%%%%%%%%%%%%%%%%%%%%%%%%%%%%%%
\subsection{Fabrication and Hallucination Risks}
AI models, including large language models (LLMs), learn from the data on which they are trained. When this data contains historical, societal, or sampling biases, these systems not only replicate but often amplify such biases in their generated outputs. This phenomenon contributes to what is commonly referred to as ``hallucination” in LLMs—the generation of plausible-sounding but factually incorrect or entirely fabricated content~\cite{bang2023multitask, maleki2024ai, yao2023llm, banerjee2024llms, lyon2023fake}.

A prominent example occurred in 2023, when a legal brief drafted using ChatGPT included fictitious case citations, ultimately leading to disciplinary action against the attorneys involved~\cite{lyon2023fake}. In scientific contexts, similar hallucinations—such as fabricated studies, inaccurate references, or unsupported claims—pose a serious risk to the integrity of scholarly literature, particularly if undetected during peer review~\cite{LLMHallucinationsZhang2025, ChatGPTHallucinatesBuchanan2024, HallucinationSysReviewChelli2024}.

Recent research highlights the vulnerabilities of LLMs to adversarial prompts and so-called ``jailbreaking” techniques, which can elicit harmful or misleading outputs that deviate from expected behavior~\cite{yi2024jailbreakLLMs}. These risks underscore the importance of rigorous validation and oversight in the use of generative AI for knowledge generation.

To ensure ethical use of AI in research and publishing, it is essential that all AI-generated content undergo thorough fact-checking and that human oversight is maintained throughout the entire research and writing process. Reliance on AI must never replace the critical judgment, domain expertise, and ethical responsibility of human researchers.

%%%%%%%%%%%%%%%%%%%%%%%%%%%%%%%%%%%%%%%%%%
\subsection{Human Skill}
The growing integration of AI into research workflows has sparked legitimate concerns about its long-term impact on researchers' skills and practices~\cite{lira2025learningcheatingaiassistance}. A central question is whether increasing reliance on AI will lead to a de-skilling of human researchers—eroding their proficiency in core analytical, methodological, or experimental techniques—or, conversely, liberate them to engage more deeply in high-level conceptual thinking and interdisciplinary problem-solving.

AI should be viewed as a tool for augmenting, rather than replacing, human expertise. When used responsibly, it can enable researchers to shift their focus toward strategic reasoning, creative inquiry, and integrative synthesis~\cite{lira2025learningcheatingaiassistance}. However, this requires that researchers maintain critical oversight, actively validate AI-generated outputs, and retain the ability to explain and defend the conclusions drawn from them. Blind reliance on automated systems undermines both scientific rigor and accountability.

Moreover, the significant computational power and technical expertise required for developing and deploying advanced AI systems raise concerns about exacerbating existing global inequities in scientific research. A growing digital divide may emerge, wherein only well-resourced institutions and nations can fully leverage AI capabilities, leaving others behind. Ethical frameworks for the use of AI in research must therefore include considerations of equitable access—to tools, infrastructure, and training—ensuring that the benefits of AI are distributed broadly across the global scientific community.

\subsection{Opportunities}
As global scientific output continues to expand at an unprecedented rate—surpassing three million publications annually and adding to an archive of over 200 million papers—the research community faces a fundamental challenge: how to meaningfully access, interpret, and build upon this vast and growing body of knowledge. This challenge is not merely logistical but epistemological, raising the question of how collective scientific progress can be sustained in an age of information overload. At the same time, emerging advances in artificial intelligence (AI) offer a historic opportunity to reimagine the entire scholarly knowledge ecosystem.

AI is rapidly becoming a transformative force across all stages of the research lifecycle. From literature discovery to peer review, from knowledge synthesis to post-publication impact analysis, AI-enabled systems promise to augment the capabilities of human researchers, streamline workflows, and expand access to knowledge. When strategically developed and ethically integrated, AI can accelerate discovery, enhance transparency, and democratize scientific participation.

One of the most immediate opportunities lies in leveraging AI for scholarly knowledge extraction and organization. With sophisticated techniques in natural language processing, machine learning, and knowledge representation, AI can distill insights from unstructured scientific texts, map complex relationships between concepts, and build dynamic, queryable knowledge graphs that evolve with the literature~\cite{JiomekongWWWConf2024, jiomekongORKG2024}. These systems can help researchers identify relevant works, surface interdisciplinary connections, and synthesize prior findings—functions that are increasingly difficult to perform manually.

The rise of generative AI, particularly large language models (LLMs), presents additional opportunities. These models are capable of summarizing research, generating hypotheses, proposing experimental designs, and even drafting initial research narratives~\cite{bolanos2024ALiteReviews}. When used responsibly and transparently, such tools can reduce barriers to participation and allow researchers to focus on conceptual innovation, interdisciplinary collaboration, and socially relevant problem-solving~\cite{lira2025learningcheatingaiassistance}.

AI also holds transformative potential for enhancing research integrity and reproducibility. Through automated verification of claims, identification of citation inconsistencies, and replication of computational workflows, AI can support a more trustworthy and transparent scholarly record. Furthermore, AI systems can assist peer reviewers and editors in assessing the factual coherence, methodological rigor, and novelty of submitted manuscripts—thereby strengthening quality control in high-volume publishing environments.

Beyond technical contributions, AI offers strategic pathways toward greater equity in global research. If developed with accessibility and openness in mind, AI-powered tools can help under-resourced institutions, early-career researchers, and non-native English speakers engage more effectively with the scientific literature. Multilingual search, simplified navigation, and context-sensitive recommendations can reduce structural barriers and foster a more inclusive research community. However, realizing this vision requires deliberate investment in equitable infrastructure, governance, and capacity-building.

\subsection{Challenges}
Despite the transformative potential of artificial intelligence in reshaping scholarly communication, its integration introduces a range of critical challenges that must be addressed with urgency and care. Without proactive intervention, these issues could undermine the integrity, equity, and sustainability of scientific research. Specifically, we highlight the following challenge areas: 

\textbf{Transparency and Reproducibility} are a primary concern for AI systems, particularly those relying on large language models and deep learning architectures. These systems often operate as opaque ``black boxes,” making it difficult to trace how conclusions are derived or to reproduce the same results. In the context of scholarly publishing—where transparency is foundational—this lack of interpretability poses a direct threat to scientific rigor.

One example of the reproducibility challenge is the distinction between whether a contribution can truly be considered `open-source' or just `open-weight'~\cite{OSI}. Open-source AI systems provide a sufficiently adequate description of the necessary resources to train a model from experimental hyperparameters, training regime, compute infrastructure and of particular importance, training data. Image generation models~\cite{Rombach_2022_CVPR, podell2023sdxl} are generally closer to open-source in this regard.

In contrast, open-weight refers to AI models whose pretrained checkpoints have been released to the open internet. Although one can download the weights and further fine-tune the model on specialized training data, access to the original pretraining data corpus is usually restricted. This is usually the case with LLMs such as Meta's Llama~\cite{touvron2023llama, dubey2024llama} lineup. 

\textbf{Workforce Integration.} Equally important, the growing reliance on AI in research workflows raises concerns about the potential de-skilling of human researchers. As automation takes over tasks such as literature reviews, writing assistance, and data interpretation, there is a risk that core competencies in critical thinking, methodological design, and scholarly communication may erode. Ensuring that AI augments rather than replaces human expertise will require robust educational frameworks and a renewed emphasis on human judgment, reflection, and accountability.

\textbf{Accountability} is also increasingly contested in AI-mediated research environments. As AI begins to generate not just summaries or visualizations, but actual hypotheses, analyses, and textual contributions, questions about responsibility and authorship become more pressing. While most major publishers currently prohibit listing AI tools as authors, ambiguities remain around the attribution of AI-assisted contributions. Without clear frameworks, the integrity of authorship and accountability is at risk.

\textbf{Bias} is also a troubling issue. Specifically, AI models are trained on large-scale corpora that often reflect historical inequities, cultural stereotypes, and disciplinary silos. Moreover, large language models are prone to hallucination—producing confident, yet factually incorrect or fabricated information. In a scientific setting, this can result in the introduction of false claims, spurious references, or misleading interpretations into the academic record, especially if such content is not carefully validated by human experts.

\textbf{Accessibility.} The promise of AI is also tempered by its unequal distribution. Access to advanced AI tools, infrastructure, and expertise is highly concentrated in well-funded institutions and nations. Without targeted efforts to address this imbalance, the deployment of AI may widen existing gaps in scientific capacity and global participation. Equitable access to AI technologies—through open-source tools, inclusive design, and infrastructure investment—is essential to prevent a deepening of the digital divide.

\textbf{Finally}, there is a broader structural challenge: the current academic reward system often prioritizes rapid publication, productivity, and novelty. These incentives may encourage misuse of AI tools to accelerate output at the expense of quality, transparency, and ethical responsibility. Without aligning institutional policies and publishing norms with the principles of responsible AI use, the long-term credibility of scholarly communication may be compromised.

Addressing these challenges will require interdisciplinary collaboration, inclusive dialogue, and the development of shared ethical and technical standards. Only through such collective effort can we ensure that AI strengthens rather than destabilizes the global research enterprise.

\section{Conclusion}
This paper has summarized the outcomes of the AAAI 2025 Bridge Program on AI for Scholarly Communication (AI4SC), which convened a diverse community of students, domain researchers, practitioners and AI experts to examine the rapidly evolving relationship between artificial intelligence and the production, organization, dissemination and use of scientific knowledge.

AI is profoundly reshaping the scholarly landscape. From automating literature reviews and extracting structured knowledge from unstructured texts to assisting in hypothesis generation, experimental design, and data interpretation, AI is accelerating the pace and scale at which research is conducted and consumed. Its influence extends beyond the confines of the research process itself, enabling advanced knowledge synthesis, context-aware dissemination, and evidence-informed policy making—thereby amplifying the societal impact of research and advancing progress toward the Sustainable Development Goals (SDGs).

Yet, alongside these opportunities come deep challenges. The integration of AI into scholarly communication necessitates a rethinking of how we assess research quality, originality, and ethical responsibility. Traditional notions of scientific rigor must now be paired with emerging requirements for transparency, verifiability, fairness, and human accountability in AI-assisted research workflows.

As AI systems become more capable, the human role in research must evolve—not in opposition to automation, but in concert with it. Researchers must maintain critical oversight, validate AI outputs, and ensure that AI augments rather than diminishes scholarly judgment, creativity, and integrity. 

\bibliographystyle{splncs04}
\bibliography{references}
\end{document}